# Enhanced ULF radiation observed by DEMETER two months around the strong 2010 Haiti earthquake

M. Athanasiou[1,2], G. Anagnostopoulos[3], A. Iliopoulos[3], G.Pavlos[3], K. David [2]

[1] {Dept. of Information & Communications, Technical University of Serres, Greece}

[2] {Dept. of Mechanical Engineering, Technical University of Serres, Greece}

[3] {Dept. of Electrical & Computer Engineering, Democritus University of Thrace, Greece}

Correspondence to: M.A. Athanasiou (mathanas@ee.duth.gr)

## Abstract

In this paper we study the energy of ULF electromagnetic waves that have been recorded by the satellite DEMETER, during its passing over Haiti before and after a destructive earthquake. This earthquake occurred on 12/1/2010, at geographic Latitude $18.46^o$ and Longitude $287.47^o$, with Magnitude 7.0 R. Specifically, we are focusing on the variations of energy of Ez-electric field component concerning a time period of 100 days before and 50 days after the strong earthquake. In order to study these variations, we developed a novel method that can be divided in two stages: first we filter the signal keeping only the very low frequencies and afterwards we eliminate its trend using techniques of Singular Spectrum Analysis, combined with a third-degree polynomial filter. As it is shown, a significant increase in energy is observed for the time interval of 30 days before the strong earthquake. This result clearly indicates that the change in the energy of ULF electromagnetic waves could be related to strong precursory earthquake phenomena. Moreover, changes in energy were also observed 25 days after the strong earthquake associated with strong aftershock activity. Finally, we present results concerning the comparison in changes in Energy during night and day passes of the satellite over Haiti, which showed differences in the mean energy values, but similar results as far as the rate of energy change is concerned.

## 1 Introduction

Earthquakes (EQs) are complex phenomena generated by rock deformation in the brittle outer part of the Earth and associated with large unpredictability, due to inherent extreme randomness (Kagan, 2007). However, in the last decades there is growing evidence that EQs precursory phenomena are detected. This evidence was based on studies of certain effects related to magnetic and telluric fields, ionospheric perturbations, nightglow observations and generation of electromagnetic (EM) emissions from DC to high frequency (HF) range and radiation belt precipitation in the upper ionosphere (Bhattacharya et al., 2007; Anagnostopoulos et al., 2010; Sidiropoulos et al., 2010). Theoretical studies and laboratory experiments suggest two main mechanisms for the production of precursor earthquake waves, namely the electromechanical mechanism and the acoustic mechanism. These mechanisms are mainly based on the deformation of rocks under pressure and temperature conditions existing in the brittle seismogenic crust, which destabilise the mechanical and electrical properties of the solids. In particular, according to the electromechanical mechanism, electric charges are generated as the result of friction and piezoelectric phenomenon that change the Earth's electric field and generates EM waves, which are considered to propagate to the upper atmosphere and ionosphere (Cress et al., 1987; Enomoto and Hashimoto, 1990; Lokner et al., 1983; Parrot et al., 1993). On the other hand according to the acoustic mechanism, gravity waves are generated before and after the earthquake. These waves propagate in the atmosphere, where their amplitudes are increased relatively to height, due to the air's density decrease, disturb the ionosphere and cause VLF emissions of electromagnetic waves ( Davies et al., 1965; Gokhberg et al., 1982; Ralchovski et al., 1985).

Most papers have been devoted to studies of ELF / VLF waves (Parrot, 1983; Gokhberg et al., 1983; Larkina et al., 1983, 1989; Parrot et al., 1985, 1989; Chmyrev et al., 1989; Serebryakova et al., 1992; Henderson et al., 1993; Xuemin Zhang et al., 2009; Akhoondzadeh et al., 2010;) and the analysis is made mostly in frequency domain (Fourier analysis), due to the large amount of data. However, Fourier analysis cannot capture some essential characteristics which are revealed in time domain analysis.

Here, we attempt an investigation of ULF satellite EM signals observed in the upper ionosphere. Our methodology have the following advantages: (1) time domain analysis of EM ULF signals detected by a satellite in the upper ionosphere is a new tool in the relative literature, (2) ELF / VLF waves compared to ULF waves are faster weakening in the ionosphere, so a ULF study by a satellite in the upper ionosphere may have directly access to the EQ preparation zone, and therefore, may give more clear results concerning EQ preparation processes (Chmyrev et al., 1989), (3) There are in general a limited number of studies on the possible relation of ULF EM waves in the upper ionosphere with earthquakes (Fraser – Smith et al., 1990; Molchanov et al., 1992). For these reasons we chose to focus on space based ULF EM emissions to study their energy changes during a long period before and after a strong EQ. Our study was based on the analysis of measurements from the DEMETER satellite around the deadly earthquake of Haiti on January 12, 2010.

## 2  Data analysis and results

For the estimation of ULF signals we used data derived from DEMETER's satellite data base. Generally, the microsatellite DEMETER was launched on June 29, 2004, its orbit altitude is 710 km, and it takes 14 orbits per day around the Earth. Among the Scientific Objectives of the DEMETER mission is the investigation of the Earth Ionosphere, disturbances due to seismic and volcanic activities. The ICE instrument allows the measurements of the three components of the electromagnetic wave field from DC up to 3.5 MHz (Berthelier et al., 2006).

In this paragraph, we present results concerning the analysis of the z-component of electric field of the ULF waves, within frequency range of 0-20Hz. It should be noted that the analysis of the other components $E_x$, $E_y$, gives similar results (not shown in this paper). In particular, the data cover a time period of 150 days, 100 before and 50 after the earthquake, corresponding to 374 semi-orbits of DEMETER satellite. These semi-orbits were carefully selected in order to be strongly related to the area where the earthquake took place. 207 of them correspond to night-passing (Up Orbits) over the Haiti, while the rest to day-passing (Down Orbits). The sampling frequency of data is 40 Hz, while the number of data per orbit is about 82000.

Figure 1 shows the waveform of $E_z$ electric field component (left) concerning the orbit 29550_1 (right) on 2010/01/09 for the time interval 01:53:30 - 02:26:00 UT. Moreover, on the x-axis of left image the values of Latitude, Longitude, L-value and Altitude of DEMETER are depicted. As it is shown within the red dashed line, there exists a significant fluctuation in the waveform, as the satellite passing over Haiti, in 02:16:00 - 02:23:00 UT and at $80^o$-$280^o$ Lat.

In Figure 2 we present a comparison of two different waveforms of $E_z$ electric field component, one (Fig. 2a) corresponding to the waveform presented to Fig.1 and the other (Fig. 2b) corresponding to a time period where no earthquake occurred in the broad seismic region of Haiti. From the observation of Figure 2b it is clear that there is no significant variation in the waveform corresponding to the period of seismic quiescence. Therefore, we assume that the anomaly in the waveform (Fig.2a) which we are interested in, can be attributed to a precursory earthquake signal.

In order to estimate the energy of the possible pre-Earthquake signal we focus on the signal shown in Figure 3a, which corresponds to the perturbed waveform (Fig. 2a). On this signal a low pass frequency filter is applied, keeping frequencies lower than 5 Hz, in order to estimate thoroughly the mean energy value. Consecutively, we use methods of Singular Spectrum Analysis (Athanasiu and Pavlos, 2001) and polynomial fitting of third order in order to remove the signal trend that corresponds to exogenous factors. The resulting signal appears in the Figure 3b. We consider that this waveform is the clear pre-Earthquake signal recorded by DEMETER satellite. The plot presented in the Figure 2c is the square power of pre-Earthquake signal and consists an estimation of its energy. In this case the mean value of energy was found to be 0.37 $(mV/m)^2$.

Figure 4a shows the mean value of energy for signals corresponding to data of 135 perturbed and unperturbed semi-orbits which cover a time interval of 100 days before the strong Earthquake, using the same procedure described previously in Fig. 3. These orbits were recorded by DEMETER satellite during its night-passing (Up Orbits) over Haiti for geographic Latitude 18.46 ± 10° ($8^0$-$28^0$) degrees and Longitude 287.47±15° (272° – 302°). This part of Earth can be considered as the seismic region around the Haiti. Also in the seismic region and for this time interval, no earthquake with magnitude greater than 5 took place. Thus, we assume that the observed signals can be related to precursor phenomena concerning the strong earthquake that occurred 100 days later. As we can see in this Figure many significant increases in the mean value of the energy are observed in a time period of one month before the main Earthquake, while the first strong signal is detected 33 days before the event.

In order to further highlight these increases in the mean energy, we estimate the mean value of energy of pre-earthquake signals per 25 days as it shown in Figure 4b. We observe that for the time interval 50-100 days before the main event, the mean value of energy takes low values around 0.09 $(mV/m)^2$. The first significant change of energy is observed for the time interval of 25 -50 days before the earthquake, where the mean value of energy is 0.13 (mV/m)$^2$, corresponding to an increase of 40%. Finally, the most significant change of energy is observed for the time interval 0-25 days before the earthquake, where the mean value of energy attains values around 0.21 (mV/m)$^2$, corresponding to an increase of 220%.

Figure 5a is similar to Figure 4a, but in this case we have rejected the values which are smaller than 0.1 (mV/m)$^2$. These low values of energy could correspond to weak signals that are not related to the seismic area of Haiti but in other external factors. As it is observed most of the values of energy are concentrated in the time interval of one month before of the earthquake. On the contrary, as far as the time interval 50-100 days before the main event is concerned, few and low-energy pre-earthquake signals are observed.

Figure 5b is similar to Figure 4b, but in this case in order to estimate the mean value of energy per 25 days we have replaced the values of energy that are greater than 0.1 (mV/m)$^2$ with zero, as it explained in the previous section. As it is shown in this figure the mean value of energy obtains low values, around 0.05 (mV/m)$^2$, for the time interval if 50-100 days before of earthquake. The first significant change of energy is observed for the time interval 25 -50 days before the main event, where the mean value of energy is 0.1 (mV/m)$^2$, an increase of 100%. Finally the most significant change of energy is observed for the time interval of 0 -25 days before the earthquake, where the mean value of energy is 0.18 (mV/m)$^2$, corresponding to an increase of 360%.

Figure 6 presents results of the mean value of energy per 25 days for the pre-earthquake signals that are recorded by the DEMETER satellite during night (Fig. 6a) and day passing (Fig. 6b) over Haiti. Figure 6a is the same as Figure 4a and has already been described. Figure 6b shows the mean value of energy for the signals recorded by DEMETER satellite during its day passing (Down Orbits) over Haiti for Latitude 18.46 ± 5$^o$ (13$^o$-23$^o$) degrees and Longitude 287.47±15$^o$ (272$^o$ – 302$^o$). The data correspond to 131 semi-orbits covering a time interval of 100 days before the Earthquake.

The comparison of Figures 6a, 6b shows that the energy perturbation of pre-seismic day-passing signals is much smaller than night-passing ones. As it is found for the time interval of 100 days before the earthquake the mean value of energy during the day-passing is 0.005 (mV/m)$^2$ while for the night-passing is 0.13 (mV/m)$^2$, namely 26 times greater. We think that this difference could be due to the strong ionization of the Ionosphere during day time that causes great attenuation in the pre-seismic signals. However, the results in Figures 6a and 6b are very similar as far as the rate of energy change is concerned. In both figures, the energy of the pre-earthquake signals increases at the same rate regarding the time before the earthquake. This result indicates the efficiency of the applied method since it can reveal precursory phenomena in data concerning the day orbits, although the energy changes are very weak.

Figure 7a represents the average energy of the pre-earthquake signals recorded by the satellite during night-passing, for 100 days before the main earthquake as well as for the aftershock signals for 50 days. As we can see, there is a significant increase in the mean energy recorded a month before the main earthquake which remains at the same levels approximately for 25 days after, while consecutively decreases gradually. In Figure 7b the overall picture of the energy change before and after the earthquake is shown. This figure shows the average energy per 25 days of the observed signal during the night, for 100 days before and 50 days after the earthquake. During the 70 days time interval after the main shock 5 strong aftershocks of magnitude greater than 5, occurred in the broad seismogenic area of Haiti. We observe that the drop in the signal's mean energy during the first 25 days after the earthquake is insignificant. On the contrary, during the period of 25 - 50 days after the earthquake there is a significant reduction of 43% compared with the maximum value. One possible explanation for this reduction of mean energy after the earthquake is a respective decrease in seismic activity in the region of Haiti after the main earthquake.

## 3 Conclusions

In this study we are focused on changes of the energy for the $E_z$ component of the electric field of electromagnetic ULF waves (0-20 Hz) that were recorded by satellite DEMETER concerning a time period of 100 days before and 50 days after a strong earthquake which took place in Haiti in 12/1/2010. For these signals we applied a novel method of two stages: first, the signals are filtered keeping only the very low frequencies and consecutively their trend was eliminated by applying techniques of SSA, combined with a third-degree polynomial filter. The results reveal a significant increase of the energy of ULF waves, up to 360%, for a period of one month before the main earthquake compared with the energy of the background. Also a gradual reduction of wave energy occurs one month after the main earthquake. Additionally, the comparison of pre-seismic day-passing and night-passing signals showed differences in the mean energy values, but similar results as far as the rate of energy change is concerned. The results of this paper clearly indicate that ULF electromagnetic waves can be very useful in revealing possible precursor seismic phenomena in Ionosphere. Following this line of evidence, other strong earthquakes occurred in low geographic latitude will be studied using ULF waveforms in order to further establish the hypothesis of ULF seismo-electromagnetic precursory emission.

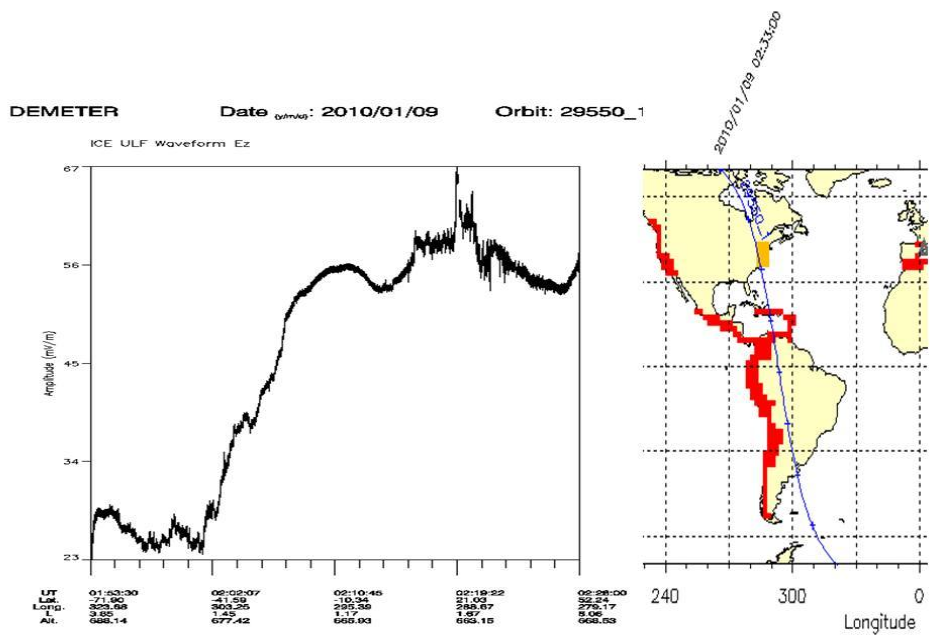

Figure 1

Figure 1. The waveform of Ez electric field component (left) concerning the DEMETER's orbit 29550_1 (right) on 2010/01/09 for the time interval 01:53:30 - 02:26:00 UT. Moreover, on the x-axis of left image the values of Latitude, Longitude, L-value and Altitude of DEMETER are depicted.

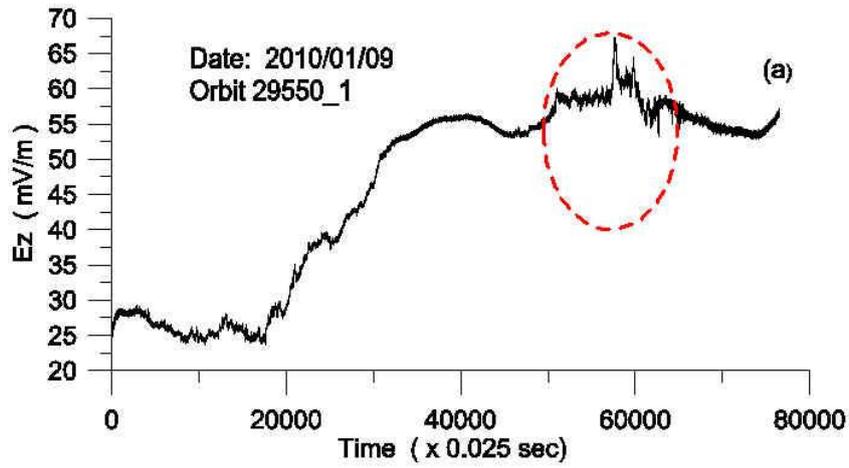

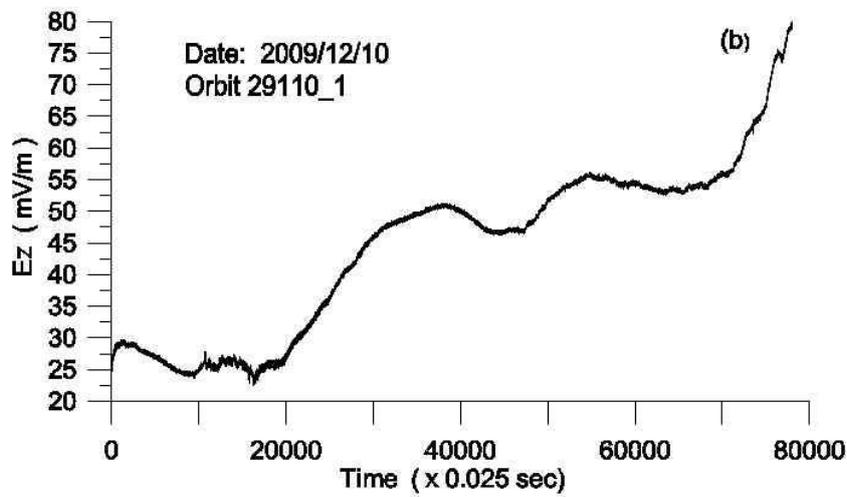

Figure 2. Comparison of two different waveforms of Ez electric field component, one (Fig. 2a) corresponding to the waveform presented to Fig.1 and the other (Fig. 2b) corresponding to a time period where no earthquake occurred in the broad seismic region of Haiti.

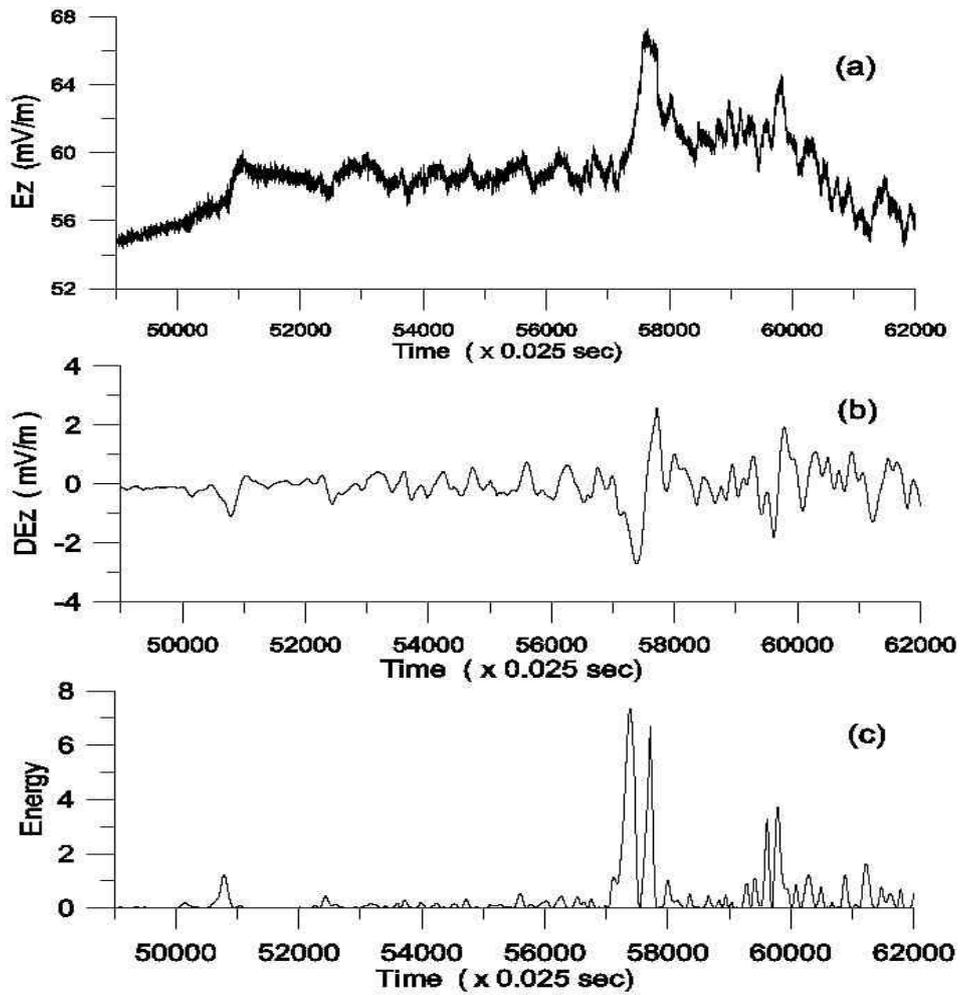

Figure 3. a) The perturbed waveform (shown in Fig. 2a) focused on time period between 49000-62000 (x 0.025 sec). b) The filtered focused perturbed signal (Fig. 3a). For filtering we used Singular Spectrum Analysis, combined with a third-degree polynomial filter. c) The square power of pre-Earthquake signal shown in Fig. 3b.

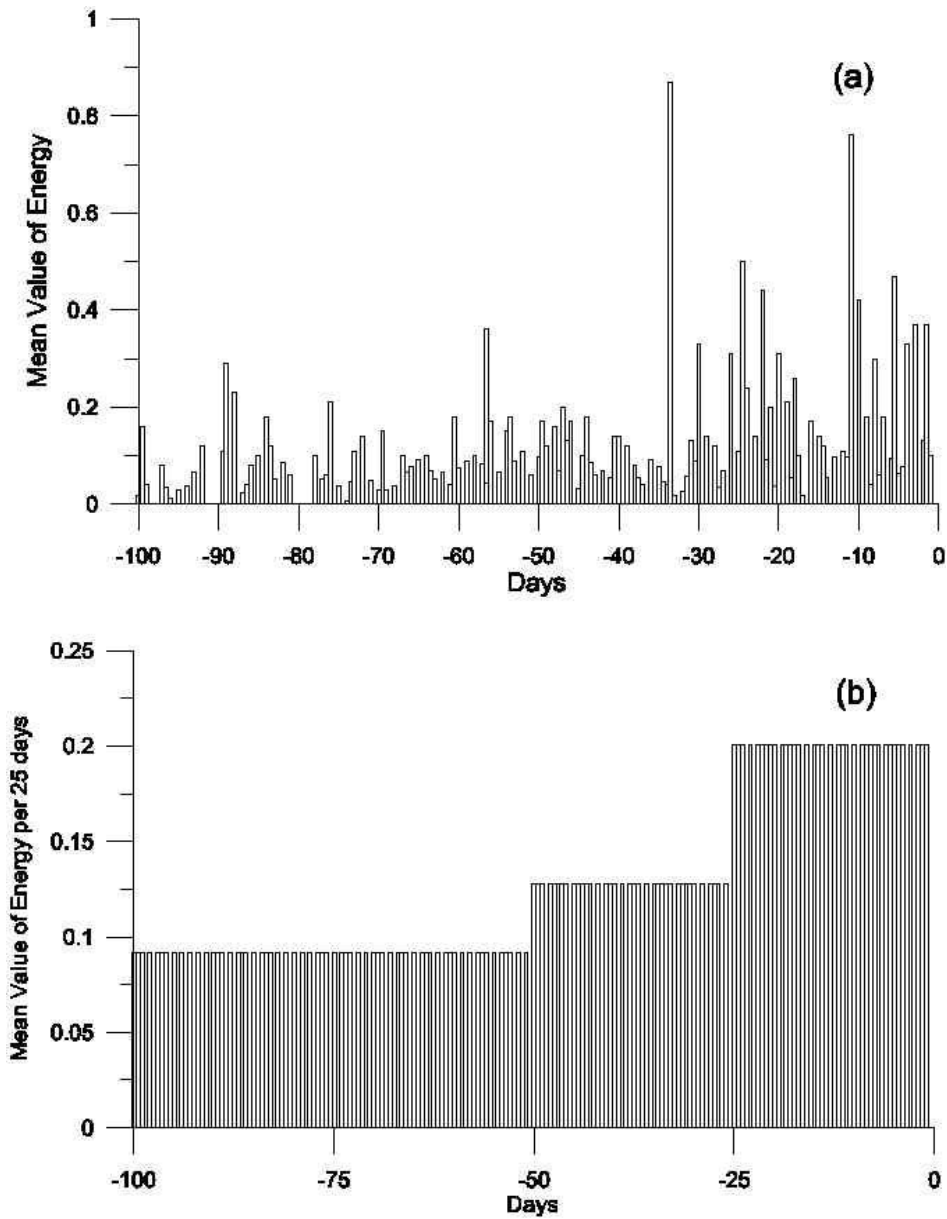

Figure 4. a) The mean value of energy for signals corresponding to data which were recorded by DEMETER satellite during its night-passing covering a time interval of 100 days before the strong Earthquake. As we can see in this Figure many significant increases in the mean value of the energy are observed in a time period of one month before the main Earthquake. b) The mean value of energy of pre-earthquake signals per 25. The most significant change of energy is observed for the time interval 0-25 days before the earthquake.

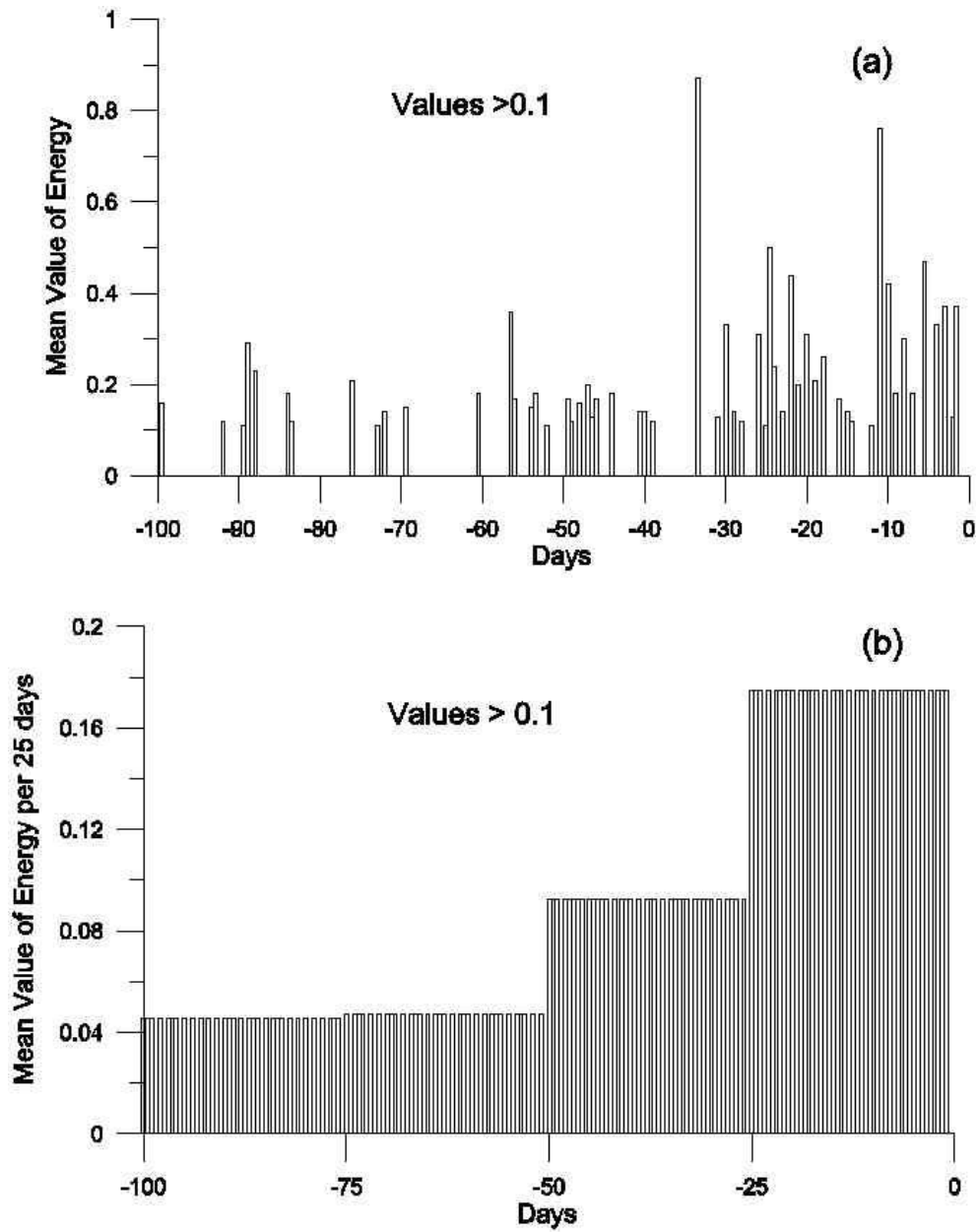

Figure 5. a) Figure 5a is similar to Figure 4a, but in this case we have rejected the values which are smaller than 0.1 $(mV/m)^2$. As it is observed most of the values of energy are concentrated in the time interval of one month before of the earthquake. b) Figure 5b is similar to Figure 4b, but in this case we have replaced the values of energy that are greater than 0.1 $(mV/m)^2$ with zero. As it is shown in this figure the most significant change of energy is observed for the time interval of 0 -25 days before the earthquake.

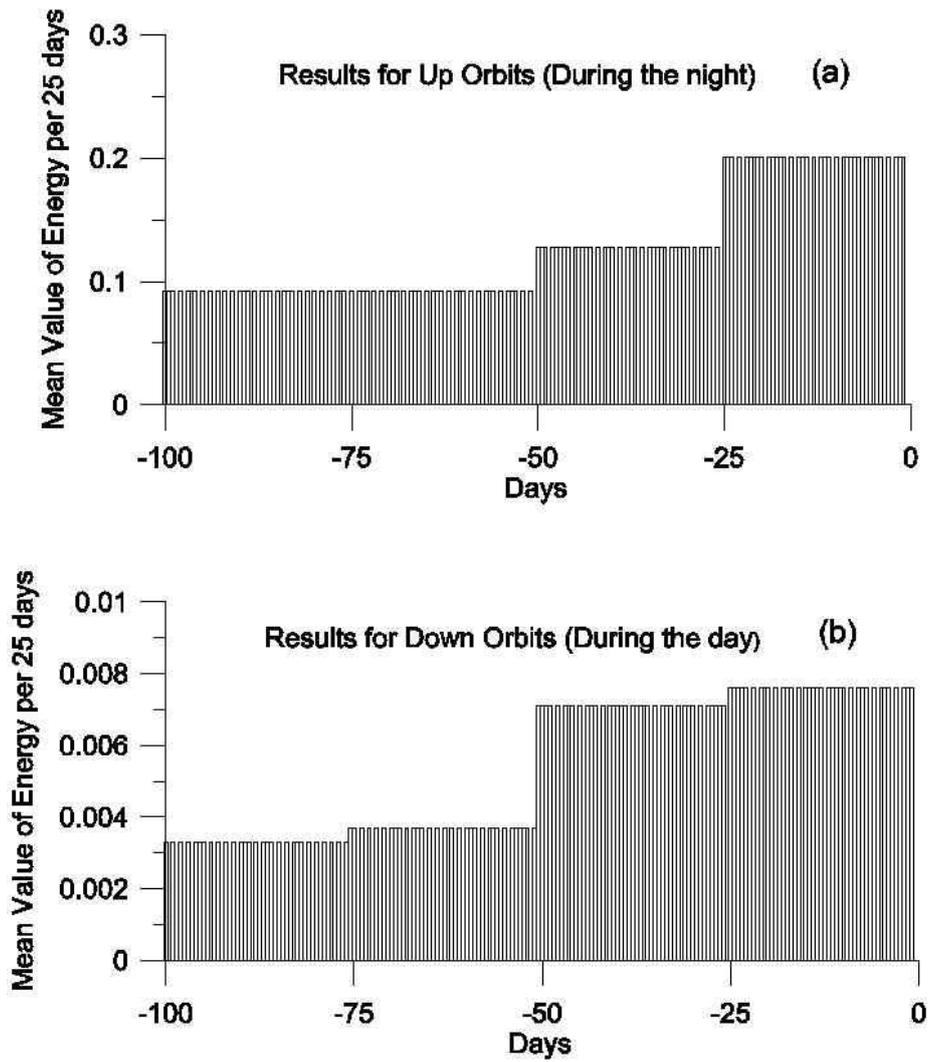

Figure 6. Results of the mean value of energy per 25 days for the pre-earthquake signals that are recorded by the DEMETER satellite during night (Fig. 6a) and day passing (Fig. 6b) over Haiti. In both figures, the energy of the pre-earthquake signals increases at the same rate regarding the time before the earthquake.

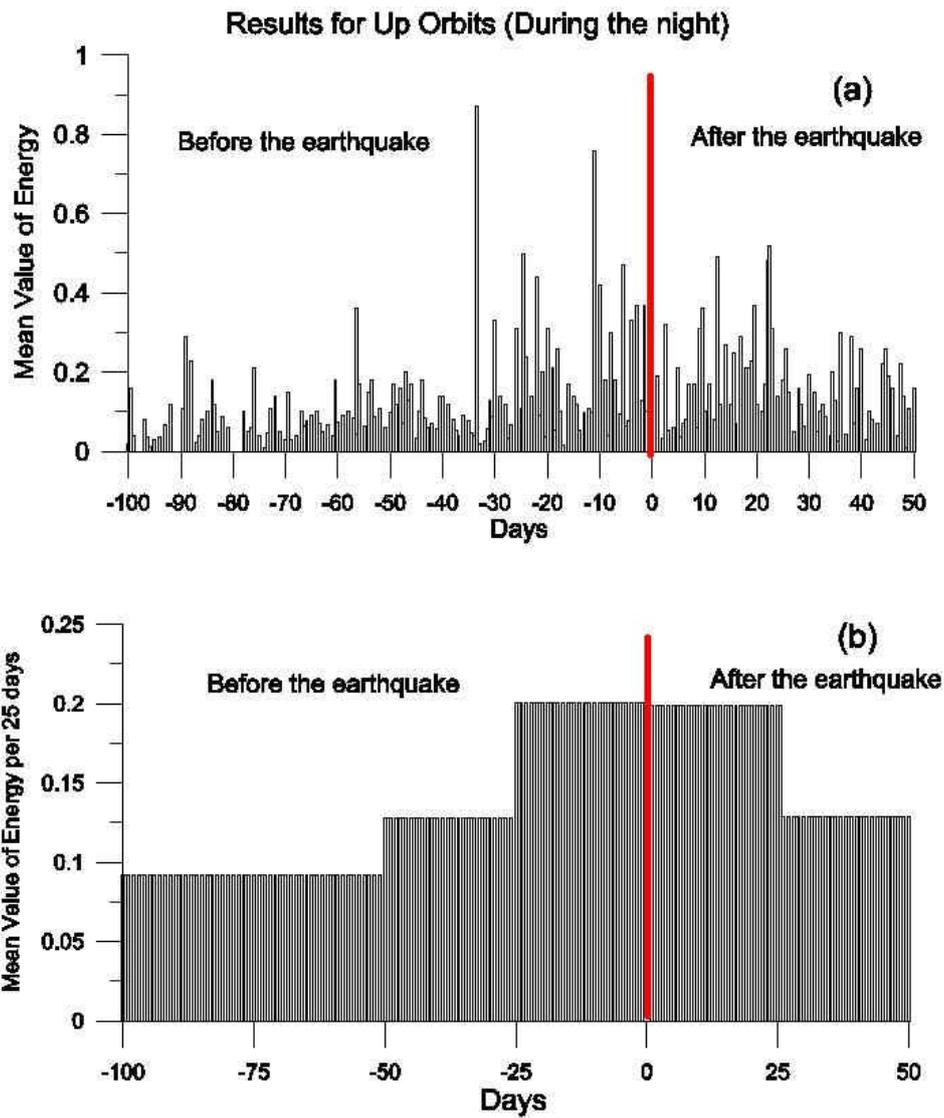

Figure 7. a) The average energy of the pre-earthquake signals recorded by the satellite during night-passing, for 100 days before the main earthquake as well as for the aftershock signals for 50 days. b) The average energy per 25 days of the observed signal during the night, for 100 days before and 50 days after the earthquake.